\documentstyle[aps,prb,psfig]{revtex}
\begin{document}
\title{Mesoscopic Superconducting Disc with Short--Range Columnar Defects.}
\author{Gregory M. Braverman, Sergey A. Gredeskul and Yshai Avishai}
\address{Ben-Gurion University of the Negev, Beer-Sheva, Israel}
\date{}
\maketitle
\begin{abstract}
Short--range columnar defects essentially influence the magnetic 
properties of a mesoscopic superconducting disc. They help the
penetration 
of vortices into the sample, thereby decrease the sample 
magnetization and reduce its upper critical field. 
Even the presence of weak defects split a giant vortex state (usually appearing 
in a clean 
disc in the vicinity of the transition to a normal state)
into a    
number of vortices with smaller topological charges. In a disc with a 
sufficient number of strong enough defects vortices are always placed 
onto defects. The presence of defects lead to the appearance of additional 
magnetization jumps related to the redistribution of vortices which 
are already present on the defects and  not to the penetration of  
new vortices. 
\end{abstract}
\begin{center}
\end{center}
\section{Introduction}
Advances in microtechnology have allowed the fabrication of
Hall probes of micron size. They were successfully applied for time-- and
space--resolved detection of individual vortices in superconductors
\cite{Hprob1,Hprob2,Hprob3,Hprob4}. Recently Geim {\it et.
al.} \cite{Geim1} developed Hall probe techniques by  
employing submicron ballistic probes of this type for studying individual
submicron samples. The use of Hall probes in the regime of ballistic
electron transport and samples of size smaller than the probe size
allowed them to make a link between the detected signal and the sample
magnetization. The experiments showed that the sample undergoes a sequence
of phase transitions of the first kind, which manifeststhemselves by
mesoscopic jumps of the magnetization curve\cite{Geim2}. These jumps 
are due to
penetrations of additional vortices inside the superconductor as the 
applied magnetic field increases. (Due to the small size of the 
sample, 
each vortex carries a magnetic flux smaller than a single 
superconducting flux quantum
$\Phi_0$.)\\ 

The results obtained in Ref.\cite{Geim2} stimulated a series of
theoretical 
works\cite{deo,schweigert,palacios,deo2,akker}. 
Deo {\it et. al.} \cite{deo,schweigert,deo2} numerically solved the 
$3D$ non-linear Ginzburg-Landau equations together with the Maxwell equations. 
They emphasized role of 
finite sample thickness and showed that S-N transition in mesoscopic disc 
could be first or second order. They also analysed the conditions of 
multi-vortex states or a giant vortex state formation, constructed a vortex 
phase diagram and expalined the experimental results\cite{Geim2}. 
Palacios\cite{palacios} considered the same problem within a variational 
approach, obtained the magnetization jumps related to the penetration of 
new vortices into the sample and showed that below 
the upper critical field for an infinite sample $H_{c2}$ 
 the vortices occupy spatially separated positions
(a vortex glass structure) while
above this field they 
always form a giant vortex located at the disc center. Choice of the 
Ginzburg-Landau parameter $\kappa=3$ resulted in a good
agreement with experimental results \cite{Geim2}. Recently Akkermans and 
Mallik\cite{akker} considered a finite sample at the dual point 
with Ginzburg-Landau parameter 
$\kappa=1/\sqrt{2}$ and obtained the magnetization curve which also came 
to the qualitative agreement with the numerical results of Ref. \cite{deo} 
and the experimental curve\cite{Geim2}. \\
 
Introduction of strong pinning centers such as columnar defects,
which can be produced by heavy--ion irradiation \cite{civale2}, 
essentially influuences the magnetic properties of the sample. 
 In bulk superconductors 
these 
defects lead to important change of the reversible 
magnetization\cite{Beek}. Even small concentration of defects 
modifies the magnetization curve of a conventional 
superconductor near $H_{c2}$ leading to a sequence of reentering 
transitions related to the two possible types of the local symmetry near 
each defect\cite{we}.\\ 

Columnar defects should also essentially change the 
magnetic properties of mesoscopic superconductors. Such defects are known 
to be insulating inhomogeneities. Generally they can be described as local 
inclusions with lower critical temperature. In the case when  
the number of defects is of the order of the number of vortices one can 
expect that they will 
essentially suppress  the magnetic response of the sample and reduce
the upper critical field $H_{c3}$. If the number of defects is
larger than the number of vortices and the defects are strong enough it 
seems plausible that all vortices could be pinned by defects. 
As the applied field changes the
vortices can change their position on the defects. These rearrangements 
should lead to 
increasing of the number of mesoscopic jumps of the magnetization curve as
compared with that of a clean sample. In the present paper we show that 
all these phenomena really take place in small enough superconducting
discs.\\

The content of the paper is as follows. In the second section we 
formulate the problem. Section \ref{sigma_sec} has an auxuliary character - 
here we reproduce some numerical results which should be used 
later on. In section \ref{potential} we describe the variational 
approach 
for the thermodynamic potential. Properties of the clean disc are 
discussed in section \ref{clean}. The main results concerning 
the disc with defects are presented in section 
\ref{defect} and summarized in section \ref{summ}.\\

For convenience any length appearing below is measured in units of the 
temperature dependent 
coherence length $\xi(T).$ In these units the penetration 
length coincides with the Ginzburg-Landau parameter $\kappa.$\\

\section{The Model.}
\indent
Consider a type II superconducting disc with thickness $d$ and radius 
$r_{0}$ containing columnar defects of size $l.$   The sample is subject 
to an applied magnetic field, which is
parallel both to the defects and to the disc axis. In what follows we 
use the dimensionless variables measuring magnetic field and vector potential 
in units of $H_{c2}=\Phi_{0}/2\pi\xi^{2}(T),$ and 
$\Phi_{0}/2\pi\xi(T)$ respectively ($\Phi_{0}$ is the 
superconducting flux quantum). Then the density of the thermodynamic 
potential and the order parameter will be measured in 
units $\alpha_{0}^{2}/\beta$ and 
$\sqrt{-\alpha_{0}/\beta}$ where $\alpha_{0}$ and $\beta$ are the standard 
Ginzburg-Landau coefficients of the clean disc. In the presence of 
defects the coefficient $\alpha$ should be modified and depends on 
coordinates
\begin{eqnarray*}
	\alpha ({\bf r})=\alpha_{0}(1-\delta\alpha ({\bf r})).
\end{eqnarray*}
 The last term in parenthesis is simply related to the 
critical temperature change $\delta T_c({\bf r})$ caused by defects:
\begin{equation}
\delta \alpha({\bf r})=\frac{\delta T_c({\bf r})}{T_c-T},
\label{alpha1}
\end{equation}
where $T_c$ is the critical temperature of a clean sample.\\

We assume that the 
disc is thin and small $d\ll r_{0}<\kappa.$ 
All the dimensions of such a disc are smaller than the penetration
depth $\kappa.$ Therefore the problem becomes essentially 2D one, 
and, moreover, it 
is possible neglect the spatial variation of the magnetic
induction ${\bf b}$ inside the disc and replace it by 
its average value $\langle{\bf b}\rangle$ \cite{palacios} (here and 
below the 
brackets $\langle ..\rangle$ mean averaging over the sample area).
As a result one gets the following expression for the 
thermodynamic potential density
\begin{equation}
G=\left\langle
-|\Psi|^2+\frac{1}{2}|\Psi|^4+|{\bf D}_-\Psi|^2+
\delta \alpha({\bf r})|\Psi|^2\right\rangle
+\kappa^2(\langle b\rangle-h)^2.
\label{PGLTP}
\end{equation}
The gauge invariant gradient ${\bf D}_-$ is
given by
$$
{\bf D}_-\equiv -i\frac{\partial}{\partial{\bf r}}+{\bf a},
$$
where the symmetric gauge ${\bf a}=
\langle b\rangle r/2\overrightarrow{\vartheta}$
is adopted.\\

According to the general approach of the Ginzburg-Landau theory one has to 
minimize the
thermodynamic potential density (\ref{PGLTP}) with respect to the
order parameter $\Psi$ with an average induction  $\langle b\rangle$ fixed 
and then to minimize the result once more with respect to 
$\langle b\rangle$. 
The first step results in 
a nonlinear differential equation with a boundary
condition
\begin{equation}
{\bf D}_-\Psi|_{r=r_0}=0,
\label{boundc}
\end{equation} 
the solution of which is
rather difficult even in the absence of defects. Therefore we 
use the variational procedure choosing the trial function as a  
linear 
combination of the eigenfunctions of the operator $({\bf D}_-)^2$
with the boundary condition (\ref{boundc}). The corresponding 
eigenfunctions $\Delta_{n,m}$ and eigenvalues $\sigma_{n,m}$ depend on the 
disc radius $r_0$. Here $m$ is an orbital number 
and 
$n$ stands for the number of the Landau level which this eigenvalue 
belongs to when the disc radius $r_{0}$ tends to infinity.
In strong enough 
magnetic field one can take into account only $n=0$ states and therefore 
the quantum 
number $n$ will be omitted in what follows. For an infinite sample 
such an approximation corresponds to neglection of higher Landau levels 
contribution which is justified from the fields $h=0.5$\cite{Brandt}. In our 
case it is adequate when the strength of defects $\delta\alpha({\bf r})$ is much
smaller than the distance between the $n=0$ and $n=1$ eigenvalues.  
Then, to 
describe states with a fixed number $N_v$ of vortices the maximal
orbital number or topological charge which  
enters the trial function should be equal to  $N_v$. Finally our 
trial function can be written as   
\begin{equation}
\Psi = \sum_{m=0}^{N_v}C_m\exp(-im\vartheta)\Delta_m,
\label{mtf}
\end{equation}
where $\Delta_m$ is given by
\begin{equation}
\Delta_m=\sqrt{\langle b\rangle}\exp\left( 
-\frac{r^2}{2}\langle b\rangle\right)
\Phi\left(\frac{\langle b\rangle-\sigma_m}{2\langle b\rangle},m+1;
\frac{r^2}{2}\langle b\rangle\right).
\label{mtf1}
\end{equation}
In Eq.(\ref{mtf}) the expansion coefficients $C_m$ serve as variation
parameters and $\Phi(a,c;x)$ in Eq.(\ref{mtf1}) is the confluent 
hypergeometric function\cite{GRy}.\\

To proceed the problem one should substitute the trial function (\ref{mtf})
into the expression (\ref{PGLTP}) for the thermodynamic potential density 
and first minimize it with a respect to the expansion
coefficients $C_m$ at an average induction
$\langle b\rangle$ fixed. As a result one obtains a system of a finite 
number of nonlinear equations for the coefficients $C_m$. This system is a finite version 
of the Ovchinnikov equations\cite{Ovch}. However in the presence of disordered 
set of defects the solution of these equations is very complicated. The 
point is that now no selection rule (successfully used in the homogeneous 
case\cite{Ovch,we,palacios}) can be applied. Thus the problem needs 
another approach.\\

In what follows we consider a disc which contains $N_d$ short--range
defects $l\ll 1$ placed at the points ${\bf r}_1,$ ${\bf r}_2,$ ..., 
${\bf r}_{N_d}.$ 
The number of defects $N_d$ is assumed to be 
larger than the maximal
possible number of 
vortices $N_v.$ As we could see  (see section \ref{clean} below) a small 
enough clean disc can accumulate vortices only 
in its center. The defects attract the 
vortices and due to their short range 
can pin the latters exactly on the positions of the  defects. Therefore 
we consider only some special configurations of vortices such that 
they
occupy only the positions of defects and the disc center. 
This choice of trial function implies the following procedure. Let us  
fix a defect configuration $\{{\bf r}_j\},$ 
$j=0,1,...,N_d,$ 
${\bf r}_0= 0,$ 
 a set of corresponding topological
 charges $\{p(j)\},$ an external magnetic field $h$ and an average induction 
 $\langle b \rangle$. Each topological charge 
 $n_j$ is non negative integer and the set  $\{p(j)\}$ satisfies the 
 condition 
 \begin{equation}
 	\sum_{j=0}^{N_d}p(j) = N_v.
 	\label{topo}
 \end{equation}
Thus our procedure accounts for the existence of  multiple vortices located on 
 the disc center or on any defect position as well. 
The trial function 
(\ref{mtf}) has zeros only at points  $\{r_j\}$ with 
miltiplicities   $p(j).$ 
The latter condition completely defines all 
coefficients $\{C_m\}$ ($m=0,1,...,N_v-1$) up to a common multiplier $C_{N_v}$, 
which we term as the order parameter amplitude. 
Further, we need to minimize the thermodynamic
potential with respect to this amplitude and the average induction.
The result has to be compared with those obtained for 
different total numbers of vortices and different sets of 
``occupation numbers''  $\{p(j)\}.$ Comparing the obtained value of the 
thermodynamic potential with that corresponding to a normal state one 
finally finds the preferable state of the disc for a fixed value of
external 
magnetic field. Repeating this procedure for various values of the magnetic 
field one could describe magnetic properties of the sample in a wide 
range of the fields up to the upper critical field $H_{c3}$. The next 
four sections are devoted to the realization of thed procedure 
described above  
and to the presentation of its results.\\

\section{Spectrum of the Operator $({\bf D}_-)^2$.}
\label{sigma_sec}
\indent
To construct the trial function (\ref{mtf}) one should first obtain the
eigenvalues 
and eigenfunctions of the 
operator  $({\bf D}_-)^2.$ This is a textbook problem and it was 
solved many times but we need the solution for various
disc 
radii and various average induction values. The 
eigenvalue equation reads:
\begin{equation}
\frac{1}{r} \frac{\partial}{\partial r} \left(
r\frac{\partial \Psi}{\partial r}\right)+
\frac{1}{r^2}\left(
\frac{\partial}{\partial \vartheta}+\frac{i}{2}\langle 
b\rangle r^2\right)^2\Psi=
-\sigma\Psi,
\label{LGLEQSG}
\end{equation}
\begin{equation}
\left.\frac{\partial \Psi}{\partial r}\right|_{r=r_0}=0.
\label{bound2}
\end{equation}
Solution of this differential equation can be 
written as:
$$
\Psi (\zeta)=Ce^{-\zeta /2-im\vartheta}\zeta^{|m|/2}
\Phi\left(\frac{|m|-m+1}{2}-\frac{\sigma}{2\langle 
b\rangle},|m|+1;\zeta\right),
$$
where $\zeta=\langle b\rangle r^2/2.$ The boundary condition (\ref{bound2})
implies the following eigenvalue equation for the quantities 
$\sigma_{n,m}$:
\begin{equation}
\frac{|m|-\zeta_{0}}{\zeta_{0}}\Phi\left(
\frac{\langle b\rangle -\sigma}{2\langle b\rangle},
|m|+1;\zeta_{0}\right)= 
\frac{\sigma-\langle b\rangle}{\langle b\rangle (|m|+1)}
\Phi 
\left(\frac{3\langle b\rangle -\sigma}{2\langle
b\rangle},|m|+2;\zeta_{0}\right),
\label{boundary}
\end{equation}
where $\zeta_{0}=\langle b\rangle r_{0}^2/2$ and the index $n$ stands for
the number of a Landau level, to which the
quantity $\sigma_{n,m}$ tends as the disc radius tends to infinity:
$$
\lim_{r_0\rightarrow\infty}\sigma_{n,m}=
\langle b\rangle (2n+|m|-m+1).
$$
 
\begin{figure}
  \centerline{\psfig{file=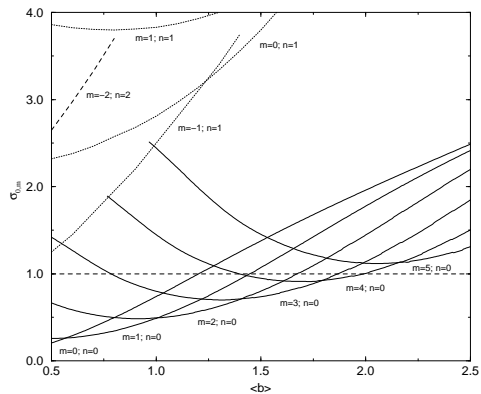,width=7.5cm,angle=270}}
  \caption{Eigenvalues $\sigma_{n,m}$ for the disc of radius
  $r_{0}=2.6$ as a function of the applied field $h$.}
  \label{sigma}
\end{figure}

We solved equation (\ref{boundary})
numerically tabulating some needed eigenvalues $\sigma_{n,m}$ 
and the corresponding 
eigenfunctions $\Delta_m$ for various quantum
numbers $n=0,1, m=1,2,3,4,5$ and disc radius $r=2.6$. The eigenvalues as 
functions of an average 
induction are shown in the fig. \ref{sigma}. These results are 
completely consistent with e.g.those obtained earlier in Ref.
\cite{benoist}.    
One can observe that the distance between the
zeroth and the first Landau levels is of the order of unity. So we can indeed
neglect in expansion (\ref{mtf}) the contributions of higher 
``Landau
levels'' as long as defects are not extremely 
strong, $\delta\alpha({\bf r})< 1$.\\

The results shown in fig. \ref{sigma} help us estimate how many 
vortices can enter the sample. Indeed, for $\sigma =1$ the eigenvalue 
equation (\ref{LGLEQSG}) coincides with the linearized 
Ginzburg-Landau equation. Therefore the maximal average 
induction $\langle b \rangle_{m}$ corresponding to $\sigma_{m}=1$ 
can be treated as the upper critical field 
 for a given orbital number $m.$ The highest of these fields is the genuine  
 upper critical field $h_{c3}$ and the corresponding value of $m$ 
 gives the topological charge of the giant vortex usually appearing in the 
 vicinity of the clean disc phase transition point (see Refs. 
  \cite{giav,schweigert} and section \ref{clean} below). 
  In the case
 $r=2.6$ considered here the highest possible field at which 
 superconductivity still exists is
$h_{c3}\approx 1.98$. This corresponds to the intersection point of the curve 
$\sigma_{4}$ and the dashed line $\sigma =1$. Thus a clean superconducting disc
of this radius at the phase transition point can accumulate only four
vortices since the curve for $n=0, m=5$ never reaches the line 
$\sigma=1$.\\  
\section{The Thermodynamic Potential}
\label{potential}
\indent
Substituting the test function (\ref{mtf}) for the order parameter into
the expression for the thermodynamic potential density (\ref{PGLTP}) one 
obtains:
\begin{equation}
G=
-\sum_{m=0}^{N_v}|C_m|^2\left(1-\sigma_m\right)I_m
+\frac{\langle b\rangle}{2}\sum_{k,m,n=0}^{N_v}
C^{*}_{m}C^{*}_{n}C_kC_{m+n-k}J_{m,n,k}+
\langle\delta\alpha|\Psi|^2 \rangle+
\kappa^2(\langle b\rangle -h)^2,
\label{PGLTP1}
\end{equation}
where the brackets $\langle ..\rangle$ mean averaging over the sample area,
$I_m\equiv\langle\Delta_{m}^{2}\rangle$, $J_{m,n,k}\equiv\langle
\Delta_m\Delta_n\Delta_k\Delta_{m+n-k}\rangle$ and
$\sigma_{m}\equiv\sigma_{0,m}$.
For the state characterized by a topological charge 
$N_{v}$ the coefficient $C_{N_{v}}$ necessarily differs from zero. We choose it 
as an amplitude of the order parameter and introduce new expansion 
coefficients $D_{m}$ and new order parameter $\psi$
\begin{equation}
\left\{
\begin{array}{ll}
C_m=C_{N_{v}} D_m,\\ 
D_{N_v}=1,\\
\Psi=C_{N_{v}}\psi.
\end{array}
\right.
\label{newdefs}
\end{equation}
Rewriting the thermodynamic potential (\ref{PGLTP1}) in terms of these new
variables and varying it with respect to the amplitude $C_{N_{v}}$ we
obtain the following expression for its extremal value:
\begin{equation}
|C_{N_{v}}|^2=\frac{\displaystyle\sum_{m=0}^{N_v}(1-\sigma_m)I_m-
\langle\delta \alpha|\psi|^{2}\rangle}
{\langle b\rangle
\displaystyle\sum_{k,m,n=0}^{N_v}D_{m}^{*}D_{n}^{*}D_kD_{m+n-k}J_{m,n,k}}.
\label{totamp}
\end{equation}	

\indent The expansion coefficients
of the
order
parameter (\ref{mtf}), (\ref{newdefs}) are completely defined by the
position
of vortices
on the defects. Let us choose some configuration of vortices 
$\{{\bf r}_j\}$.
In this set there are points occupied by a single
vortex ($p(j)=1$) and points corresponding to multiple vortices with
topological charge $p(j)>1.$ Then the set of coefficients
$\{D_m\}=\{\varphi^{-1} C_m\}$ can be calculated from the following 
system of
$N_v$ linear equations:
\begin{equation}
\sum_{m=0}^{N_v-1}D_m\exp (-im\vartheta_j)
\Delta_{m}^{(p(j))}({\bf r}_j)=
\exp (-iN_v\vartheta_j)\Delta_{N_v}^{(p(j))}({\bf r}_j),
\label{syst}
\end{equation}
where the notation $f^{(n)}(x)$ is used for the $n$th derivative.\\

The ``inhomogeneous term'' in Eq.(\ref{PGLTP}) which is 
proportional to $\delta\alpha ({\bf r})$ appears due to columnar 
defects. We have already mentioned that defects are supposed to be 
short-range ones. In this case this term can be represented as a sum 
over defects. For the Gaussian form of defects
\begin{equation}
\delta \alpha({\bf
r})=\frac{\alpha_{1}}{l^{2}}\sum_{j=1}^{N_d}\exp\left(
-\frac{({\bf r}-{\bf r}_j)^2}{2l^2}\right)
\end{equation}
the ``inhomogeneous'' term in (\ref{PGLTP1}) in the leading approximation
with respect to our small parameter $l$ 
can be rewritten as
\begin{equation}
\langle\delta\alpha|\psi|^2\rangle =
\frac{2\alpha_{1}}{r_{0}^{2}}
\sum_{j=1}^{N_d}|\psi({\bf r}_j)|^2.
\label{inhterm}
\end{equation}
Substituting 
equations (\ref{newdefs}), (\ref{totamp}) and (\ref{inhterm}) into
Eq. (\ref{PGLTP1}) we obtain the final expression for the 
thermodynamic potential of the disc with defects:
\begin{equation}
G=
-\frac{
\left(
\displaystyle\sum_{m=0}^{N_v}|D_m|^2(1-\sigma_m)I_m-
\displaystyle\frac{2\alpha_{1}}{r_{0}^{2}}
\displaystyle\sum_{j=1}^{Nd}
|\psi({\bf r}_j)|^2
\right)^2
}
{2\langle b\rangle
\displaystyle\sum_{k,m,n=0}^{N_v}D_{m}^{*}D_{n}^{*}D_kD_{m+n-k}J_{m,n,k}}
+ 
\kappa^2(\langle b\rangle -h)^{2}.
\label{BGLTP1}
\end{equation}	
\indent We solve the system
(\ref{syst}) for each combination of vortices on the defects in order to
find the set of expansion coefficients $\{D_m\}$ as a function of the 
average induction $\langle b\rangle$. The set of coefficients 
is then pluged into expression (\ref{BGLTP1}) for the
thermodynamic potential $G$ at a fixed applied field $h$. Now we
can find the average magnetic induction $\langle b\rangle$ at which the
thermodynamic potential (\ref{BGLTP1}) has a minimal value at fixed
applied field and configuration of vortices. After that we must repeat
this procedure for different configurations and different values of 
the 
applied field. 
As a result,
we obtain a number of data sets
for the thermodynamic potential as a function of the applied field 
for 
different configuration of vortices.Then for each value of an applied
field we should choose the  preferable vortex configuration which minimizes
the thermodynamic potential. This enables us to obtain the disc 
magnetization 
as a function of the applied magnetic field.\\

\section{Clean Disc}
\label{clean}
\indent
We start from the case of a clean disc with radius $r_0=2.6$ 
and $\kappa=3$. Although this value of
$\kappa$ limits the condition $\kappa \gg r_{0}$, the chosen region of
applied fields enables us to neglect the spatial variation of the magnetic
induction \cite{palacios}.The
maximal number of vortices in such a disc equals four (see  section 
\ref{sigma_sec}). Due to the sample geometry and small maximal number of 
vortices they can form only a number of symmetric configurations when 
some vortices occupy the disc center and the others are placed 
away from the center in such a way that they
form a regular polygon. All these configurations  
are presented in fig.2. In cases (b), (h); (d) and (g) 
the topological charge of the multiple vortex at the origin is equal to 
$2;3,4$ respectively. In cases (c),(e),(f),(h),(i),(j) the shifted 
vortices are place at a distance  $\rho$ from the origin.\\
\begin{figure}
  \centerline{\psfig{file=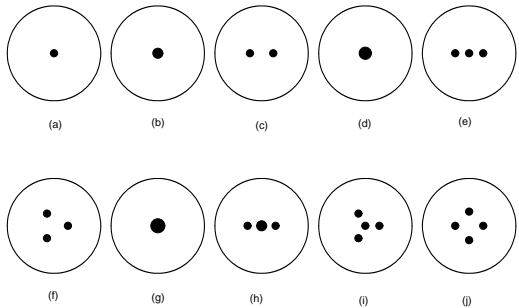,width=6.5cm,angle=270}}
  \caption{Possible configurations of vortices inside clean disc of 
radius $r_{0}=2.6$.}
  \label{conf}
\end{figure}
For a given vortex configuration the expansion coefficients
$\{D_m\}$ can be calculated from the system of linear equations
(\ref{syst}). For each possible vortex configuration we substitute these
coefficients into the expression for the thermodynamic potential of the 
clean disc
 \begin{eqnarray*}
G=
-\frac {\left(
\displaystyle\sum_{m=0}^{N_v}|D_m|^2(1-\sigma_m)I_m
\right)^2}
{2\langle b\rangle
\displaystyle\sum_{k,m,n=0}^{N_v}D_{m}^{*}D_{n}^{*}D_kD_{m+n-k}J_{m,n,k}}
+ 
\kappa^2(\langle b\rangle -h)^{2}.
\end{eqnarray*}	
and minimize it with respect to the average induction
$\langle b\rangle$. We repeat this procedure for all configurations and
for various distances of vortices from the disc center inside each
configuration. Thus the problem has three variational parameters:
the type of vortex configuration (fig.2), the distance $\rho$ of
vortices from the disc center and the average induction $\langle
b\rangle$. We changed the distance $\rho$ by step of $\delta \rho =0.1r_0.$
Numerical calculation showed that because of
the disc small size only configurations in which $\rho=0$ (fig.
2 (a),(b),(d),(g)) gain the energy. So within the calculation accuracy
$\delta\rho=0.26$ we have only a multiple vortex at the disc center 
with a possible topological charge $p(0)=1,2,3,4$.\\ 
\begin{figure}
  \centerline{\psfig{file=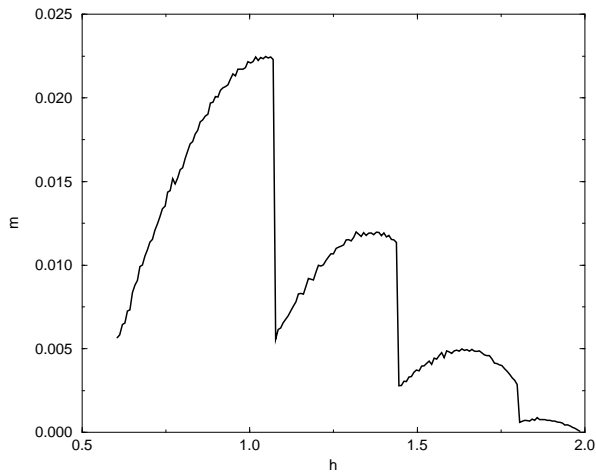,width=9cm,angle=270}}
  \caption{ Magnetization curve of a clean superconducting disc of
radius $r_{0}=2.6$.
  }
  \label{enhom}
\end{figure}

The dimensionless magnetization $m=h-\langle b\rangle$ of the clean 
disc is presented in fig. \ref{enhom}. Penetration of an
additional vortex
inside the sample is manifested by  magnetization jump.
 Each branch of the curve corresponds to the one-, two-, three-
and four--vortex states. This result is similar to that
obtained by Palacios \cite{palacios} and Deo {\it et. al.} \cite {deo} for
discs  with larger radii and it will be used in th next section devoted to 
the magnetic properties of the disc with defects.\\  

\section{Disc with Defects}
\label{defect}
\begin{figure}
  \centerline{\psfig{file=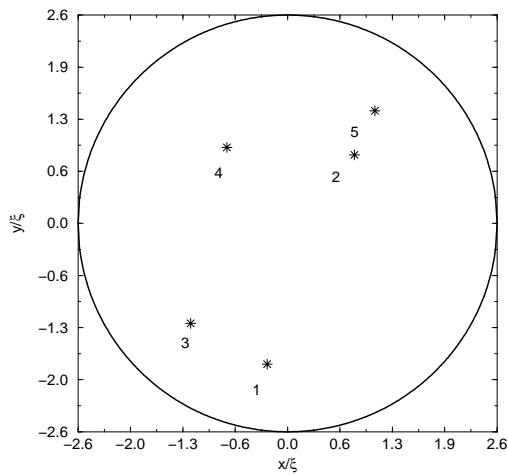,width=9cm,angle=270}}
  \caption{ Defects positions in the disc
  }
  \label{defects}
\end{figure}

\indent
In the case of disc with defects, 
one should take into account the defects configuration and 
minimize the thermodynamic potential (\ref{BGLTP1}). 
We present below the results for a single 
configuration of the defects obtained with the help of a random number 
generator. 
We hope that it is rather typical (see fig. \ref{defects}). 
In any case the results obtained below for 
this 
configuration enable us to demonstrate all the new features characterizing 
the magnetic properties of a sample with defects and to confirm all the  
expectations formulated above in the Introduction.
The coordinates of defects 
are collected in Table
\ref{defs}. (Note that all distances are measured in the temperature 
dependent coherehce length units.)\\

We analyze the thermodynamic properties of the disc for
various values of defect strength $\alpha_{1}$.
This constant can be easily varied experimentally by changing the
sample temperature (see Eq.(\ref{alpha1})). To present the results more 
clearly 
we 
collect all configurations of vortices which will be realized for values 
considered for the defect strength in Table 
\ref{vconfg}.\\
 
\begin{figure}
  \centerline{\psfig{file=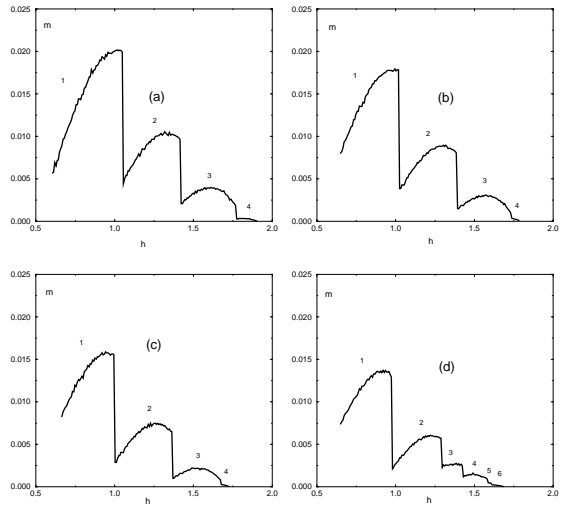,width=8cm,angle=270}}
  \vspace{0.5cm}
  \caption{ Magnetization curve of the superconducting disc of radius
$r_{0}=2.6$ and $\kappa=3$ in the presence of defects with effective
coupling constants $\alpha_{1}=0.04$ (a), $\alpha_{1}=0.08$ (b),
$\alpha_{1}=0.12$ (c) and $\alpha_{1}=0.16$ (d).
  }
  \label{engs1} 
\end{figure}

The left column of the table contains the values of the coupling 
constants.
The upper line enumerates the vortex configurations 
ordered with accordance to their appearance with the growth of a 
magnetic field. The same numbers enumerate
different regions of the 
magnetization curves on figs. \ref{engs1}, \ref{engs3.1}. Note that 
the last 
configuration in each line appears just before the phase transition to the 
normal state at the upper critical field $h_{c3}$. Then, each configuration is 
described by an ordered sequence of six numbers. The $j$-th number is 
equal to the topological charge located at the point ${\bf r}_{j-1}$. 
In  
other words the first number is the topological charge at the disc center, 
the second number is the topological charge at the first defect and so on. 
For example configuration $\{211000\}$ corresponds to double vortex at the 
disc center and two single  
vortices placed at the first and the second defects.\\
\begin{table}[h!]
\caption{Coordinates of Defects.}
\centering
\begin{tabular}
{||l|l|l|l||}\hline
  & $x$ & $y$ & $r$\\\hline
1 & $-0.253$ & $-1.755$ & $1.773$\\\hline
2 & $0.830$ & $0.856$ & $1.192$\\\hline
3 & $-1.205$ & $-1.248$ & $1.734$\\\hline
4 & $-0.755$ & $0.948$ & $1.212$\\\hline
5 & $1.083$ & $1.405$ & $1.774$\\\hline
\end{tabular}
\label{defs}
\end{table}
\begin{table}[h!]
\caption{Configurations of vortices.}
\centering
\begin{tabular}{||l||l|l|l|l|l|l||}\hline
     & $1$ & $2$ & $3$ & $4$ & $5$ & $6$\\\hline
0.04 & $100000$ & $200000$ & $300000$ & $40000$ & $$ & $$\\\hline
0.08 & $100000$ & $200000$ & $300000$ & $310000$ & $$ & $$\\\hline
0.12 & $100000$ & $200000$ & $300000$ & $211000$ & $$ & $$\\\hline
0.16 & $100000$ & $200000$ & $101100$ & $300000$ & $101200$ & 
$211000$\\\hline
0.3  & $000101$ & $000110$ & $000111$ & $001110$ & $000130$ &
$001210$\\\hline
\end{tabular} 
\label{vconfg}
\end{table}

\begin{figure}
  \centerline{\psfig{file=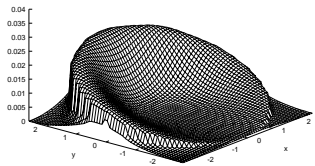,width=5cm,angle=270}}
  \caption{ Square modulus of the order parameter for $\alpha_{1}=0.08$ 
  at an applied field $h=1.753.$ The vortex configuration is $\{310000\}.$ 
}
  \label{ord1}
\end{figure}

\begin{figure}   
  \centerline{\psfig{file=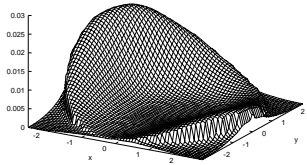,width=5cm,angle=270}}
  \caption{ Square modulus of the order parameter for
$\alpha_{1}=0.12$ at an applied field $h=1.7.$ The vortex configuration 
is $\{211000\}.$ 
  }
  \label{ord2}
\end{figure}

\begin{figure} 
  \centerline{\psfig{file=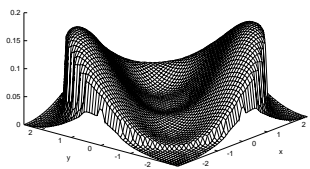,width=5cm,angle=270}}
  \caption{ Square modulus of the order parameter for
$\alpha_{eff}=0.16$ at an applied field $h=1.4$ 
The vortex configuration is $\{101100\}.$ 
  }
  \label{ord3} 
\end{figure}

\begin{figure}
  \centerline{\psfig{file=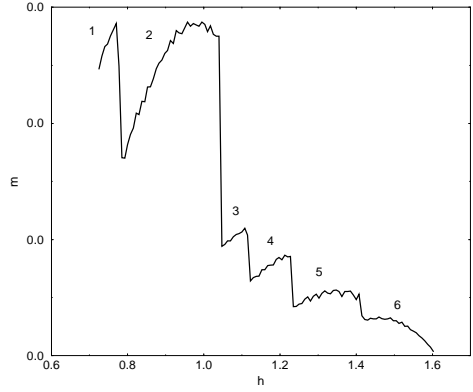,width=7.5cm,angle=270}}
  \caption{ Magnetization curve of the superconducting disc of radius
$r_{0}=2.6$ and $\kappa=3$ in the presence of defects with an effective
coupling constant $\alpha_{1}=0.3.$
  } 
\label{engs3.1}
\end{figure}

We start from small values of the defect strength. 
The corresponding magnetization curves are shown in fig. \ref{engs1}.
The first part (a) of this figure describes the magnetization curve for 
a 
sample with $\alpha_{1}=0.04.$ 
 Because of the small value of the coupling constant, this part is 
 qualitatively equivalent to that for a clean disc. Each branch of the 
 magnetization
curve corresponds to a one--, two-- , three-- and four--vortex states. 
These branches are divided by jumps of the magnetization
which are caused by penetration of an additional vortex inside the sample.
However, even in this case 
some new features caused by defects are manifested. We particularly 
refer to the  
suppression of magnetization, penetration of new
vortices at lower fields and decreasing of the upper critical field in
comparison with the results for the clean sample (see fig.
\ref{enhom}). 
Magnetization of the samples with $\alpha_{1}=0.08$
(fig. \ref{engs1}.b) and with $\alpha_{1}=0.12$ (fig. \ref{engs1}.c)
have the same number of mesoscopic jumps as in the previous case. This
means that all the jumps are still due to vortex penetrations. However
a new interesting feature appears near the phase transition point. The
four-multiple vortex at the disc center is split. In the case 
 $\alpha_{1}=0.08$ (fig. \ref{engs1}.b.4) three-multiple vortex remains 
 at the center and one 
 more vortex occupies the first defect (configuration $\{310000\}$). 
 The corresponding distrubution of the absolute value 
 square of order parameter is presented in fig. \ref{ord1}.
More complicated splitting is observed in the case $\alpha_{1}=0.12$ 
(fig. \ref{engs1}.c.4) 
Two vortices remain at the disc center, one occupies the first
defect and another one occupies the second defect (configuration $\{211000\}$).
The square modulus of the order parameter is plotted in fig.  \ref{ord2}.\\
In the two latter cases the defect strength was relatively 
small. Therefore the defects could partially destroy the 
giant vortex state with maximal multiplicity which precedes 
the transition 
to the normal state. 
Further
increasing of the coupling constant leads to appearance of additional
mesoscopic jumps related to the  rearrangement of the vortices on the 
defects as the applied magnetic field changes. 
Consider the case  $\alpha_{1}=0.16$ (fig. \ref{engs1}.d). 
At small values of the applied field one gets one- and two--vortex
states at the disc center.
However, when the third vortex is allowed to penetrate (fig.
\ref{engs1}.d.3) the multiple vortex is destroyed and the vortices 
occupy the disc center, the second
defect and the third defect (configuration $\{101100\}$). 
Plot of the square modulus of the order parameter
for this vortex configuration can be found in fig. \ref{ord3} (to 
present the 
plot more clearly the orientation of the axes is changed with respect to the two 
previous plots).\\

With further increasing of the applied field the
system turns again into the three-multiple vortex state at the disc center 
(fig. \ref{engs1}.d.4). So in the same sample two different vortex 
configurations with the same total topological charge are possible. 
When the fourth vortex penetrates the disc the 
three-multiple vortex state splits
again (fig. \ref{engs1}.d.5) into double vortex at  the third
defect, one vortex at the disc center and another one at the second
defect
(configuration $\{101200\}$). The 
appearance of the second vortex on the third defect is a result of a very
restricted space of the trial functions.  Indeed, according to Eq.
(\ref{BGLTP1}) any defect which is already occupied by a vortex is put out
of the game and one can not gain energy adding one more vortex to the same
defect. This means that in a wider variational space the configuration
$\{101200\}$ would be replaced by another one which should
be more preferable. At the same time it  will necessary lead 
to the corresponding magnetization jump.\\

With increasing of the applied field we have a new jump of the
magnetization curve, which is caused by rearrangement of the vortices into
the configuration $\{211000\}$ identical to that of the four vortex state 
in the case 
$\alpha_{1}=0.12$.\\ 

Thus one can see that 
the stronger defects are the greater is the tendency of vortices to occupy 
defects. The 
destruction of the giant vortex at the disc center begins near the
upper 
critical field. 
Increasing the defect strength destroys the centered multiple vortices 
with lower multiplicity. The preferable arrangement of the vortices 
corresponds to the maximal reduction of the square order parameter 
modulus.\\

\begin{figure}
  \centerline{\psfig{file=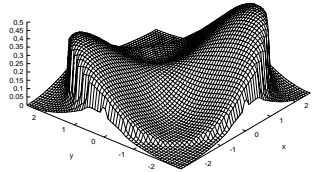,width=5cm,angle=270}}
  \caption{ Square modulus of the order parameter for
$\alpha_{1}=0.3$ at an applied field $h=0.71.$ 
The vortex configuration is $\{000101\}.$ 
  }
  \label{ord4}
\end{figure}

\begin{figure}
  \centerline{\psfig{file=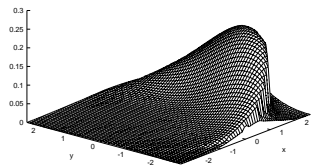,width=5cm,angle=270}}
  \caption{ Square modulus of the order parameter for
$\alpha_{1}=0.3$ at an applied field $h=1.31.$ 
The vortex configuration is $\{000130\}.$ 
  }
  \label{ord8}
\end{figure}

\begin{figure}
  \centerline{\psfig{file=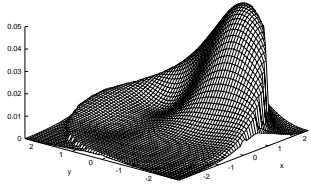,width=5cm,angle=270}}
  \caption{ Square modulus of the order parameter for
$\alpha_{1}=0.3$ at an applied field $h=1.55.$
The vortex configuration is $\{001210\}.$ 
  }
  \label{ord9}
\end{figure}

At strong coupling constant one expects to get states where all 
vortices are placed onto defects for all values of the  
applied field. Consider the results of studying the case $\alpha_{1} =
0.3.$ The magnetization curve of such disc is shown in fig. \ref{engs3.1}.
Penetration of vortices inside the disc with such strong defects
occurs at values of the applied field smaller than that of
the previously considered discs with relatively weak defects. Because of
that, already at a field $h=0.6$ the disc accumulates two vortices (fig.
\ref{engs3.1}.a,b.1). Their configuration is  $\{000101\}$ (see fig.
\ref{ord4}). 
As the applied field increases this configuration is changed by another 
one $\{000110\}$ with the same total topological charge.
Three vortices appearing at higher fields always occupy three 
different defects. The corresponding configurations are  $\{000111\}$ 
and  $\{001110\}$. Two configurations with total topological charge 
four are realized. Both 
contain a multiple vortex on one of the defects. The first configuration 
appearing in relatively low field is $\{000130\}.$ Here one has 
three-multiple vortex on the fourth defect. The second configurarion 
$\{001210\}$ preceding the transition to the normal state at $h_{c3}$ 
contains a double vortex at the third defect. 
Plots of the square modulus of the order patameter for these cases are shown 
in figs. \ref{ord8} and \ref{ord9}. Thus in the case of a strong 
defect $\alpha_{1} =0.3$ considered here the number of magnetization jumps within the 
same field region is twice the number of possible values of the total 
topological charge. We do believe that in a disc of the same radius 
containing more defects this number will increase.\\

\begin{figure}
  \centerline{\psfig{file=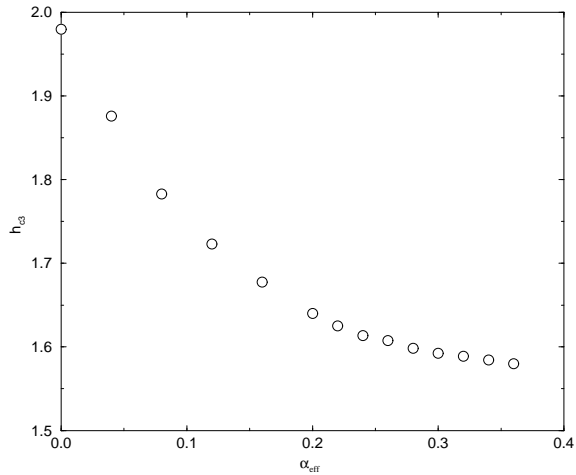,width=9cm,angle=270}}
  \caption{ The upper critical field as a function of the defect 
  strength.
  } 
\label{hc3_}
\end{figure}

We already mentioned that the presence of attractive defects reduces the 
upper critical field $h_{c3}$ at
which the thermodynamic potential of the superconductor (\ref{BGLTP1})
becomes
equal to zero (the thermodynamic potential of normal metal).
Figures \ref{engs1} and \ref{engs3.1}
show that the
larger the defect strength $\alpha_{1}$ is the lower is the
transition field. The dependence on the upper critical field of the defect 
strength   
$\alpha_{1}$  is shown in fig. \ref{hc3_}.\\

\section{Summary}
\label{summ}
\indent
Summarizing, we studied magnetic properties of mesoscopic superconducting 
discs with disordered attractive columnar defects. The number of 
defects is assumed to be larger than the maximal possible number of 
vortices accumulated by the disc. We obtained the 
magnetization 
curves for various strengths of defects in a wide region of the 
applied magnetic field. 
The results show that the defects help the penetration of vortices 
into the sample. They 
also reduce both the value of the magnetization and the upper critical field. 
Even 
the presence of weak defects can split the giant vortex state at the disc center
(usually 
existing in a clean disc of small radius) into
vortices with smaller topological charges. This splitting occurs in the 
vicinity of the upper critical field.   
Strong ehough defects always pin all vortices, splitting multiple vortex 
states at the disc center in all field region. This leads to the appearance of 
additional mesoscopic jumps in the magnetization curve related not to the 
penetration of new vortices into the sample but to redistribution of 
vortices within the set of defects. The number of these jumps enlarges 
increases with  
the number of defects.\\

\section{Acknowledgments}
This research is supported by grants from the Israel Academy of 
Science ``Mesoscopic effects in type II superconductors with
short-range pinning inhomogeneities'' (S.G.) and ``Center of Excellence'' 
(Y.A.) and by a DIP grant for German Israel collaboration (Y.A.).\\

\end{document}